\documentclass[sigconf, authorversion=true]{acmart}

\AtBeginDocument{%
	\providecommand\BibTeX{{%
			\normalfont B\kern-0.5em{\scshape i\kern-0.25em b}\kern-0.8em\TeX}}}

\setcopyright{none}

\copyrightyear{2022}
\acmYear{2022}

\acmConference[ICSSP 2022]{16th International Conference on Software and Systems Process}{May 20--22, 2022}{Pittsburgh, USA}

\usepackage{framed}
\usepackage{csquotes}

\usepackage{fancybox}

\copyrightyear{2022} 
\acmYear{2022} 
\setcopyright{rightsretained} 
\acmConference[ICSSP'22]{Proceedings of the International Conference on Software and System Processes and International Conference on Global Software Engineering}{May 20--22, 2022}{Pittsburgh, PA, USA}
\acmBooktitle{Proceedings of the International Conference on Software and System Processes and International Conference on Global Software Engineering (ICSSP'22), May 20--22, 2022, Pittsburgh, PA, USA}
\acmDOI{10.1145/3529320.3529321}
\acmISBN{978-1-4503-9674-5/22/05}

\begin{document}

\title[How Can We Develop Explainable Systems?]{How Can We Develop Explainable Systems? \\ Insights from a Literature Review and an Interview Study}

\author{Larissa Chazette}
\email{larissa.chazette@inf.uni-hannover.de}
\orcid{0000-0001-6093-8875}
\affiliation{%
  \institution{Leibniz University Hannover \\Software Engineering Group}
  \streetaddress{Welfengarten 1}
  \city{Hannover}
  \country{Germany}
  \postcode{30167}
}

\author{Jil Klünder}
\email{jil.kluender@inf.uni-hannover.de}
\orcid{0000-0001-7674-2930}
\affiliation{%
  \institution{Leibniz University Hannover \\Software Engineering Group}
  \streetaddress{Welfengarten 1}
  \city{Hannover}
  \country{Germany}
  \postcode{30167}
}

\author{Merve Balci}
\email{aylakci@stud.uni-hannover.de}
\orcid{0000-0001-6794-8400}
\affiliation{%
  \institution{Leibniz University Hannover \\Software Engineering Group}
  \streetaddress{Welfengarten 1}
  \city{Hannover}
  \country{Germany}
  \postcode{30167}
}

\author{Kurt Schneider}
\email{kurt.schneider@inf.uni-hannover.de}
\orcid{0000-0002-7456-8323}
\affiliation{%
  \institution{Leibniz University Hannover \\Software Engineering Group}
  \streetaddress{Welfengarten 1}
  \city{Hannover}
  \country{Germany}
  \postcode{30167}
}

\renewcommand{\shortauthors}{Chazette et al.}

\begin{abstract}
Quality aspects such as ethics, fairness, and transparency have been proven to be essential for trustworthy software systems. Explainability has been identified not only as a means to achieve all these three aspects in systems, but also as a way to foster users' sentiments of trust. Despite this, research has only marginally focused on the activities and practices to develop explainable systems. To close this gap, we recommend six core activities and associated practices for the development of explainable systems based on the results of a literature review and an interview study. First, we identified and summarized activities and corresponding practices in the literature. To complement these findings, we conducted interviews with 19 industry professionals who provided recommendations for the development process of explainable systems and reviewed the activities and practices based on their expertise and knowledge. We compared and combined the findings of the interviews and the literature review to recommend the activities and assess their applicability in industry. Our findings demonstrate that the activities and practices are not only feasible, but can also be integrated in different development processes.
\end{abstract}


\maketitle
\section{Introduction}
Decisions made or supported by software have an increasing influence on our daily lives. It all starts with deciding which route to take to work. Is there an accident somewhere on the route? Or more traffic than on a typical morning? Navigation software applications (apps) retrieve such kind of information to recommend the best route under the present circumstances. However, the rationale behind the software's suggestions is not always clear, potentially leading to confusion and misunderstandings. Besides that, at the same time that we are getting more dependent on those systems to make decisions, these systems are also getting more complex. As a result, there is an increasing demand for system transparency and fairness~\cite{Brun2018}, motivating the development of explainable systems.

We refer to an explainable system as a system presenting an explanation about a given system aspect to an addressee (e.g, the end user), such that the addressee can understand this aspect of the system \cite{Chazette2021}. System explainability can positively influence user experience, and can foster more trust in systems \cite{Bussone2015,Chazette2020,Eiband2018}. For example, proposing a different route accompanied by the information that there is an accident (or that, simply spoken, the new route requires less time) increases the chance that the driver follows the suggestion and also increases transparency in the communication between the system and the user.

Recent research has strongly focused on the relevance of explanations and how they should be provided in a software application (cf. \cite{Cheng2019,Langer2021,Wiegand2020}). For example, the correct amount of explanations is required to keep users as informed as necessary, but too many explanations can overwhelm them \cite{Wiegand2020}. Other authors explore the difference between stakeholders' needs with respect to explanations and how those should be presented \cite{Cheng2019,Langer2021,Martin2021}. 

However, so far, it remains unclear how explainable systems can be developed \cite{Brunotte2022}. That is, how does a software process look like that specifically supports the integration of explanations? Which phases of the software lifecycle are affected by the need for explanations? Are existing processes, activities, methods, and practices sufficient for the development of these systems? 

In this paper, we want to provide deeper insights in how to develop such systems. We combine results from a  literature review, in which we analyzed 79 papers with respect to the proposed methods and practices for developing explainable systems, with the feedback provided by 19 practitioners in an interview study. We synthesize all insights into six core activities that are crucial for the creation of explainable systems: (1) vision definition, (2) stakeholder analysis, (3) back-end analysis, (4) trade-off analysis, (5) explainability design, and (6) evaluation. These activities cover three phases of the software lifecycle: requirements engineering, design, and evaluation. All participants recommended a user-centered development approach for the development of explainable systems, confirming our previous assumption \cite{Chazette2020}. In the context of this paper, we refer to a \textit{phase} as a step in the software lifecycle; to \emph{activity} a collection of practices to achieve a goal; to \emph{practice} as a specific way of performing a task during an activity. 

\textit{Outline.} The rest of the paper is structured as follows: In Section~\ref{sec:background-and-related-work}, we present background information and related research. Section~\ref{sec:research-design} presents our research design. Section~\ref{sec:results} summarizes our results which we discuss in Section~\ref{sec:discussion}. We conclude our paper in Section~\ref{sec:conclusion}.

\section{Background and Related Work}\label{sec:background-and-related-work}
The development of explainable systems has been subject to recent research. Eiband et al.~\cite{Eiband2018} present an end-user-centered participatory process that incorporates perspectives of users, designers, and providers to develop transparent and comprehensible user interfaces, both of which are quality aspects related to explainability. The authors distinguish between three types of mental models that summarize these different perspectives and work as stages in the process of defining what needs to be explained in a system: the \textit{expert mental model}, the \textit{user mental model}, and the \textit{target mental model}. The expert mental model summarizes the key components of the algorithm in an
\enquote{optimal} version of a user mental model. The user mental model describes how the user perceives the system. Any
differences and matches with the expert mental model
are recorded to form the target mental model, which combines the essential components regarded as relevant and useful in the mental models of users and experts. 

Tsai and Brusilovsky~\cite{Tsai2021} use the framework proposed by Eiband et al.~\cite{Eiband2018} to develop explainable interfaces for a social recommender system. In a first step, the authors create an expert mental model to collect the functionality of five system models. In a second step, they develop a user mental model to ponder on design decisions for an explainable interface. They performed a survey with 14 participants to create this user mental model. In a third step, the authors develop a target mental model using an experimental setting with 15 participants. The target mental model summarizes information on the most important components that require explanations. In a fourth and last step, they develop 25 different explainable user interfaces for five recommendation models which they evaluate to select the best options.

Mohseni et al.~\cite{Mohseni2021} describe a framework for the design and evaluation of explainable systems concentrating on the aspects that need to be explained, and how and when the explanation should be presented. Their framework is based on a literature review in the area of explainable artificial intelligence. In addition, the authors propose an iterative process to develop such systems. Wolf~\cite{Wolf2019} proposes to develop explainable systems based on scenarios, as they allow to elicit and analyze requirements. The author introduces so-called \textit{explainability scenarios} to help developers focus on what end users need to comprehend in a given scenario and what type of explanation they need. 

Chazette and Schneider~\cite{Chazette2020} discuss factors that affect the inclusion of explainability as a necessary quality aspect within a system and that also affect the design choices toward its operationalization. The authors encourage the use of user-centered practices when developing explainable systems~\cite{Chazette2020}. They argue that explainability and usability are intrinsically related and, therefore, methods from usability engineering would be suitable in the development of explainable systems. Schoonderwoerd et al.~\cite{Schoonderwoerd2021} present a user-centered design approach with reusable patterns for explanations. These patterns cover domain analysis, requirements elicitation, the design of the system, and the evaluation of interactions.  Weigand et al. \cite{Weigand2021} specifically propose the user-centered development process defined in the ISO 9241-210 \cite{ISO9241-210:2019} to develop explainable systems, which takes into account user-centered principles and an iterative approach that incorporates feedback loops. 

In this paper, we review all above-mentioned papers (and several more) to provide a comprehensive and synthesized process to develop explainable systems. 

\section{Research Design}\label{sec:research-design}

Our research design follows a mixed approach, with a literature review and an interview study. In the remainder of this section, we discuss the research questions and the data analysis procedures. 

\subsection{Research Goal and Research Questions}
The overall goal of the research presented in this paper is to \textit{provide recommendations on activities and practices to support the development of explainable systems}. To reach this goal, we pose the following research questions:

\noindent
\textbf{RQ1:} \textit{Which activities and practices can be used when developing explainable systems?} This question aims to offer an overview of activities and practices that have been proved to be effective in the development of explainable systems.  

\noindent
\textbf{RQ2:} \textit{How can an explainable system be developed?} Based on the findings of RQ1, our objective is to synthesize the identified practices into activities that should be incorporated into the development process of explainable systems. The activities are evaluated based on practitioners' experiences in an interview study.

\subsection{Data Collection and Analysis}
The data collection consists of two parts: (1) a literature review and (2) an interview study. While the literature review strives for an overview of already existing approaches to develop explainable systems (RQ1), the interview study was intended to collect feedback on these existing ideas, which were synthesized in six activities (RQ2). 

\subsubsection{Literature Review}
The literature review aims to provide a broad overview of the employed methods and practices when developing explainable systems. We opted to perform a rapid literature review rather than a systematic literature review based on our goal and time restrictions. Rapid reviews are a form of knowledge synthesis in which components of the systematic review process are simplified or omitted to produce information in a timely manner~\cite{Grant2009,Tricco2015}. We performed this review based \emph{in parts} on the guidelines provided by Kitchenham et al.~\cite{kitchenham2009systematic}, but deviated to some extent given our constraints. We are aware that this reduces the generalizability of our results  compared to more thorough reviews, but we deemed this form of literature study adequate for our purposes. The threats to validity introduced by this decision are presented in Section \ref{sec:Threats}. Nevertheless, we performed all steps required to get an overview of existing literature, namely (1) definition of the search string, (2) definition of inclusion and exclusion criteria, (3) selection of the database(s), (4) definition of the termination criterion, (5) execution, and (6) data analysis.

\textit{(1) Definition of the search string.} The search string is based on a combination of the keywords we wanted to cover with this study, namely \textit{explainability, software,} and \textit{system}. Note that, as we did not expect papers to explicitly propose a process to develop such systems, we did not include words like \textit{practice,} or \textit{process} in the search string. As this decision enlarged the solution space, this is not a threat to validity. We specifically focused on the fields of requirements engineering and human-computer interaction since the development of explainable systems is highly dependent on understanding the needs of the end user. The final search string results from this line of thought and is extended by synonyms and abbreviations (e.g., \textit{SE} for \textit{software engineering}). This leads to the following search string:
\vspace{7pt}
\\
\fbox{\parbox{0.98\linewidth}{(explain* \textit{OR} XAI) \\\textit{AND} \\(system \textit{OR} software \textit{OR} design \textit{OR} interface\\
\textit{OR} HCI \textit{OR} ''human-computer interaction'' \\ \textit{OR} RE \textit{OR} ''requirements
engineering'' \\ \textit{OR} SE \textit{OR} ''software engineering'')}}
\\~\\

\textit{(2) Definition of inclusion and exclusion criteria.} We eliminated studies and publications that are not relevant for answering our research questions. To increase the objectivity of the decision on the inclusion or exclusion of a paper, we formulated the inclusion and exclusion criteria presented in Table~\ref{Tab:Criteria}. In particular, as explainability is an interdisciplinary research topic, we needed to explicitly remove publications that are not related to computer science (EC$_5$). 

\begin{table}[h!]
	\centering
	\caption{Inclusion (IC) and Exclusion (EC) Criteria}
	\label{Tab:Criteria}
	\begin{tabular}{p{0.1\columnwidth}p{0.8\columnwidth}}
		\hline
		ID & Description \\
		\hline
		$IC_1$   & The paper presents  activities or practices to develop explainable systems. \\
		$IC_2$   & The paper proposes a methodology or a process to develop explainable systems.  \\
		\hline
		$EC_1$   & The paper does not mention  activities or practices.\\
		$EC_2$   & The paper is not a peer-reviewed contribution to a conference or a journal.                                         \\
		$EC_3$   & The paper is not accessible (via university licenses).  \\
	 $EC_4$   & The paper is neither written in German nor in English. \\
		$EC_5$   & The paper is not related to computer science.\\
		\hline                                                                  
	\end{tabular}
\end{table}

\textit{(3) Selection of database(s).} In rapid reviews, sources are limited due to time constraints of searching. Since the process of selecting studies for a literature review can be very laborious and time-consuming, we opted to perform our search using Google Scholar (GS). GS retrieves results from all major databases, such as ACM Digital Libraries, IEEExplore, and Web of Science. However, the use of just one database introduces a threat to validity that we discuss in Section \ref{sec:Threats}. 

\textit{(4) Definition of the termination criterion.} According to Wolfswinkel et al. \cite{Wolfswinkel2013}, a literature review is never complete but at most saturated. This saturation is achieved when no new concepts or categories arise from the inspected data. We followed this approach to decide when to conclude our search process. It is important to note that we do not assume saturation in the traditional sense, based on all available data; rather, we assume saturation with respect to the subset of publications inspected in our literature search. More specifically, we decided to end our search as soon as we found 50 papers one after another without gaining new insights, i.e., without at least one paper extending our solution space of methods and practices. Therefore, we are confident that the number of sources we found provide sufficient insights to adequately answer our research questions. 

\textit{(5) Execution.} We inserted the search string in GS, leading to 446 publications that were selected based on their titles. The application of the execution criteria to these papers led to a removal of 367 publications: We removed 300 publications not related to the focus of our study (EC$_1$), 12 publications that were not peer-reviewed (EC$_2$), 20 that were not accessible (EC$_3$), 4 that were neither written in German nor in English (EC$_4$), and 31 publications due to the missing relation to computer science (EC$_5$). All remaining publications were peer-reviewed and meet one of the two inclusion criteria, so that we considered 79 papers as relevant for our study. 

\textit{(6) Data Analysis.} From the 79 papers found to be relevant, we extracted information that was summarized in a concept matrix.  We extracted information on the  practices, as well as on the purpose of using them during the development process. We grouped this information into categories, in a process that consisted of three steps: 1) We clustered the extracted information to avoid duplicates and to get a unique list of elements, being careful to preserve the traceability between practice and phase or its purpose during the development process. 2) Based on the number of papers that mentioned these practices (at least two), we considered them to be more or less relevant for our analysis. 3) We discussed the set of practices until we reached an agreement about the final set.

To create the activities, we analyzed the associated information about the purpose of these practices and how they are used in the development process. We grouped this information into categories, following the same three-steps process as before. Then, we classified the practices into the corresponding activities based on the information in the literature. Finally, we assigned each activity to the appropriate phase of the software lifecycle. This resulted in the first version of the six activities (and corresponding practices), based only on information from the literature. For this first version, the activities were created using the most relevant practices from this concept matrix, which can be found in 39 publications from our review (Tab. \ref{tab:processLiterature}). The activities and practices were given as a reference to the participants during the interview study.

\subsubsection{Interview Study} To get feedback on the applicability of the process resulting from the literature review, we conducted an interview study with practitioners. We elaborated an interview protocol with questions and tasks for the participants. The interviews were semi-structured and helped us to learn from the practitioners' experiences. Our goal was to combine this feedback with the first version of the six activities, synthesized from our literature review. 

\textit{Interview structure.} The interviews were exploratory and consisted of two predefined tasks and a set of predefined questions that could be adapted during the interview. The overall structure of the interviews is presented below:

\begin{enumerate}
    \item Welcome
    \item Presentation of the topic and the structure of the interview
    \item Task 1: Draw a diagram of the company's existing software development process
    \item Definition of explainability and presentation of a scenario
    \item Example: Scenario on the planned development of a system that should be explainable. The participant should develop a process allowing to address explainability in the process
    \item Task 2: Draw a diagram of a development process for explainable software systems
    \item Follow-up question: Applicability in industry
    \item Closure
\end{enumerate}

After asking the practitioners about the current software development process in their companies, an introduction to the topic of explainability was given, since we cannot assume that everyone has the same understanding of what explainability means. Then, each interviewee was asked to describe and draw the current development process in his/her company, including activities, methods, and practices. Using think-aloud, the interviewees were asked to explain their thoughts. Afterwards, we introduced explainability by using the definition provided by Chazette et al.~\cite{Chazette2021} (presented in the introduction). We highlighted the relevance of explainability by presenting a scenario: The participants were asked to put themselves in the role of a process engineer in a fictitious company that wants to develop an autonomous car. As process engineer, they should develop a process (including activities, methods, and practices) that explicitly addresses explainability for a system in a self-driving car. We asked the participants to draw the process as well, and describe their rationale. To facilitate this step, we provided a list of activities and practices from software engineering and human-computer interaction, including the ones found in our literature review. Afterwards, we asked about their opinion on the applicability of the recommended process in the industry.

\textit{Participant selection.} In this study, we wanted to interview persons who have experience in software development. That is, we considered software developers and software engineers (among others) suitable interview participants. In addition, we invited product owners and requirements engineers to participate, as the literature on the development of explainable systems put a strong focus on requirements engineering. We invited 87 practitioners via LinkedIn (43 invitations), personal contacts (35 invitations) and via one contact person in a company (11 invitations). In total, 19 experts accepted our invitation, resulting in a response rate of 22\%. The main reasons for not participating were time constraints, holidays, or workplace policy that prohibited participation.

\textit{Setting.} All interviews were performed online via BigBlueButton (hosted by the university). In case of technical problems, we switched to Jitsi Meets (also hosted by the university). All participants agreed upon recording the interview. Based on the recorded data, we transcribed all interviews. In addition, we used the collaborative tool \textit{Miro} as a virtual board for the tasks. 

\textit{Pilot interviews.} We performed three pilot interviews with PhD students, which resulted in changes to the interview and tasks' structure.

\textit{Data Collection.}
We conducted 19 interviews with participants from seven companies. The interviews had an average duration of 60 minutes and were mostly conducted in German. Two interviews were conducted in English.

A complete overview of the participants is presented in Table~\ref{Tab:demographics}. 

The participants had an average age of 32.2 years (min: 23 years, max: 41 years, SD: 4.6 years) and an average of 6.6 years of experience (min: 2 years, max: 17 years, SD: 4.4 years). Almost half of them work in small companies with less then 50 employees, five work in medium-sized companies with less then 250 employees, and five work in large companies with more than 250 employees. 

\begin{table}[h]
\centering
	\caption{Overview of the participants' demographics}
	\label{Tab:demographics}
\begin{tabular}{lllll}
\hline
&&&Years of &Company\\
ID & Role & Age & experience &  size \\
\hline
1 & Requirements engineer & 37 & 17 & small \\
2 & Product owner & 32 & 5 & small \\
3 & Requirements engineer & 27 & 2 & small \\
4 & Developer & 23 & 3 & small \\
5 & Developer & 32 & 7 & small \\
6 & Requirements engineer & 31 & 5 & small \\
7 & Developer & 27 & 2 & small \\
8 & Developer & 33 & 9 & small \\
9 & Developer & 25 & 4 & small \\
10 & Requirements engineer & 35 &8  & large \\
11 & Developer & 34 & 10 & medium \\
12 & Developer & 28 & 6 & medium \\
13 & Developer & 35 & 3 & large \\
14 & Product owner & 29 & 5 & large \\
15 & Product owner & 41 & 16 & large \\
16 & Product owner & 34 & 2 & medium \\
17 & Requirements engineer & 35 & 8 & large \\
18 & Requirements engineer & 35 & 2 & medium \\
19 & Product owner  & 39 & 12 & medium\\
\hline
\end{tabular}
\end{table}

\textit{Data Analysis.} The 19 interview transcripts resulted in almost 95K lines of text. We deductively categorized the statements based on the guidelines presented by Mayring~\cite{mayring2019qualitative} to systematically retrieve insights from qualitative data. That is, given the overall goal of our study, we considered activities, methods, and practices from human-computer interaction or software engineering as categories and classified the interview data in these categories. In addition, we analyzed the interviewees' sketches of process recommendations and compared them to our literature findings. We discussed the differences and merged the findings throughout the course of several sessions until we arrived at a final version of six core activities assigned to their corresponding phases in the software lifecycle (Fig. \ref{fig:process}).

\subsection{Validity Procedures and Threats to Validity}
\label{sec:Threats}
Both studies are subject to some threats to validity. We tried to mitigate some threats by implementing validity procedures. However, some factors still threaten the validity of our results. We describe the validity procedures and discuss the threats to validity according to the classification by Wohlin et al.~\cite{wohlin2012experimentation}.

\textit{Conclusion Validity.}
Since the results of our study are based on insights from 79 publications and a total of 19 interviews, they must not be over-interpreted. However, we are confident that the data suffices for the analysis presented in this paper, as we strive for an overview and not for the ultimate truth. Future studies are required to extend our conclusions, and to derive more thorough and fine-grained results that strengthen the reliability of our findings.

\textit{Internal Validity.}
The internal validity of the results of the interview study may be threatened by the fact that a researcher was present during the data collection, adjusting the questions on-the-fly, when necessary. This might have introduced a researcher bias, because questions can be asked in a way that influences the answer. To mitigate this threat, each research step, including the interview protocol, was reviewed by the other authors of this paper. In the literature review, in case of doubts, we included a paper rather than excluded it to avoid loss of data. 

The decision to use only Google Scholar to perform our literature search also poses a threat, even though GS covers all major databases. However, Yasin et al. \cite{Yasin2020} investigated the suitability of GS for literature reviews and found that GS was able to retrieve 96\% of primary studies. Therefore, we consider the choice of GS to be appropriate given our time constraints and since our goal was to conduct a rapid literature review rather than a systematic literature review. However, we acknowledge that this is a limitation of our study and do not dismiss the need for a systematic review to provide an even more complete set of relevant publications.

\textit{Construct Validity.} 
To mitigate the threat of construct validity, we implemented a thorough review process. While one of the authors planned and conducted the two studies, each step was carefully reviewed and discussed by other authors of this paper. In addition, for the interview study, we conducted pilot interviews that led to slight adjustments in the questions and in the interview structure. During data analysis, we carefully extracted and discussed relevant information from both publications and interview transcripts. 

The choice of interview participants is also a threat to construct validity. As there are still few development teams that have experience in the development of explainable systems, the insights we gained in the interviews are based on a hypothetical scenario. Therefore, it is possible that the hypothetical ideas that emerged from the interviews do not work in industry. Future studies are required to retrieve quantitative data from a real-world setting. Furthermore, all interviewees work in an agile work environment, biasing their view on plan-driven approaches to develop explainable systems and reducing the solution space derived from the interview study. Due to the selection of participants and the results from the literature review, our results do not equally cover all phases of the software lifecycle. Studies that explicitly provide insights into the technical part of the development of explainable systems are required.   

\textit{External Validity.}
Based on the literature review and the interviews, we derived six core activities for the development of explainable systems. These activities have only been analyzed with regard to their potential applicability in industry (but not for their \enquote{real} applicability). So far, it was not possible to conduct a concrete case study highlighting how these activities can be used to support a real development process. Nevertheless, we are confident that the conclusions made are also correct for other companies and scenarios. However, we must not assume that the activities are suitable for all possible contexts.
\section{Results}\label{sec:results}
In this section, we present the results related to both our literature review and the interviews. We give an overview of our findings rather than an in-depth analysis since the space allowed for research papers is limited to elaborate on all findings.

\subsection{Literature Review}

We conducted the literature review to assess what activities and practices are used or recommended for the development of explainable systems. We considered the unique properties of explainable systems as described in the literature in order to classify the appropriate activities and practices.

Many authors recommend the integration of a user-centered development strategy for the creation of explainable systems (cf.~\cite{Naiseh2021,Tsai2021,Cheng2019,Chazette2020,Wang2019}). An explainable system provides explanations to address knowledge gaps, which are highly specific to each individual. Furthermore, explanations constitute a communication interface between humans and machines. These are two important reasons to incorporate approaches and experiences from human-machine interaction inside the process.

 \begin{figure*}[t]
	\centering
	\includegraphics[width=0.99999\textwidth]{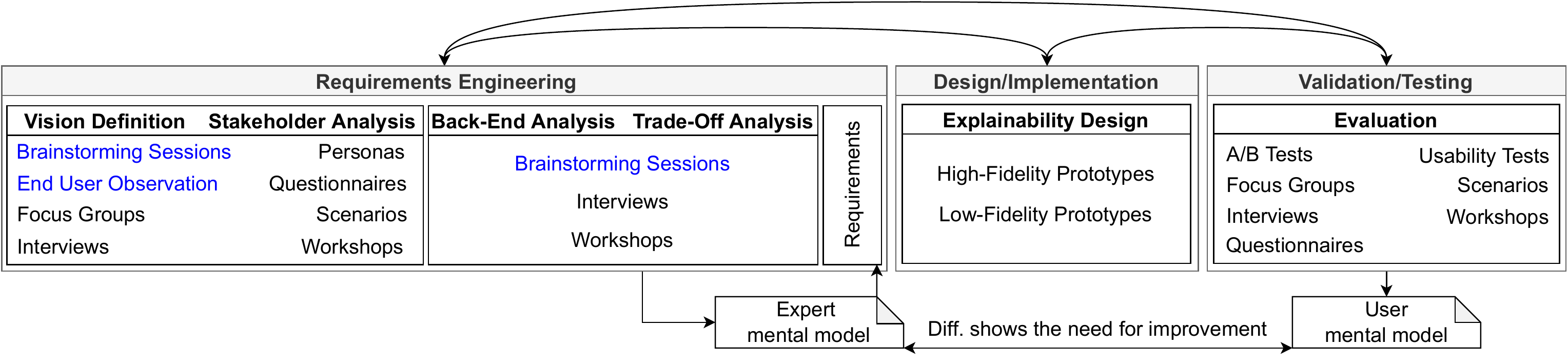}
    \caption{Six core activities and recommended practices for developing explainable systems, assigned to the respective phases of the software lifecycle}
	\label{fig:process}
\end{figure*}

We found no studies that expressly stated that a completely different methodology from existing development models is required. Instead, we identified six core activities and associated practices, presented in Figure \ref{fig:process}. The practices in blue (brainstorming sessions and end user observation) were only mentioned in the interview study, while the others were mentioned in both the literature and the interview study. We assigned the activities to the corresponding phases of the software lifecycle. Table \ref{tab:processLiterature} shows the relevant practices  for the development of explainable systems identified in the literature, organized by phase. 

The identified activities can be integrated into any software process that implements the software lifecycle, being waterfall or agile. One of the main distinctions between waterfall and agile methodologies is how the lifecycle is conducted. Agile is an incremental and iterative approach that repeats specific phases of the lifecycle; whereas waterfall is a linear and sequential approach that conducts the lifecycle once. The arrows at the top of Fig. \ref{fig:process} indicate that the activities can be iterated as needed.

\begin{table}[]
	\caption{Practices found in the literature, organized by phase}
	\label{tab:processLiterature}

\begin{tabular}{p{0.4\columnwidth}p{0.5\columnwidth}}
\toprule
\multicolumn{1}{l}{\textbf{Practice}} & \multicolumn{1}{l}{\textbf{Sources}} \\
\midrule
\multicolumn{2}{c}{Requirements Engineering} \\
\cmidrule(lr){1-2}\\

Focus Groups/Workshops & \cite{Naiseh2021},\cite{Yang2021}  \\
Interviews & \cite{Brusilovsky2019},\cite{Cech2019},\cite{Cirqueira2020},\cite{Du2019},\cite{Eiband2018},\cite{Grimaldo2019},\cite{Hall2019},\\
& \cite{Holder2021},\cite{Mohseni2021},\cite{Schneider2021},\cite{Tsai2021},\cite{Yang2021} \\
Mental Models & \cite{Eiband2018},\cite{Tsai2021}\\
Personas &  \cite{Andres2020},\cite{Barkan2020},\cite{Chazette2020},\cite{Sun2022},\cite{Jesus2021},\cite{Ramos2021},\\
&\cite{Subramonyam2021},\cite{Xu2019}\\
Questionnaires & \cite{Eiband2018},\cite{Mohseni2021},\cite{Mualla2020},\cite{Ramos2021},\cite{Tsai2021B}\\
Scenarios &  \cite{Andres2020},\cite{Cirqueira2020},\cite{Ehsan2020},\cite{Ehsan2021},\\ & \cite{Langer2021},\cite{Mualla2020},\cite{Naiseh2021},\cite{Omeiza2021},\cite{Rebanal2021},\\
&  \cite{Ribera2019},\cite{Subramonyam2021},\cite{Wolf2019},\cite{Xu2019},\cite{Sun2022}\\

\midrule
\multicolumn{2}{c}{Design/Implementation} \\
\cmidrule(lr){1-2}\\
Low-Fidelity Prototypes & \cite{Eiband2018},\cite{Kulesza2010},\cite{Subramonyam2021},\cite{Xie2020},\cite{Zhou2020}\\ 
High-Fidelity Prototypes & \cite{Brusilovsky2019},\cite{Bussone2015},\cite{Cheng2019},\cite{Eiband2018},\cite{He2015},\cite{Heier2021}, \\
&\cite{Holder2021},\cite{Mohseni2021},\cite{Naiseh2021},\cite{Schoonderwoerd2021},\cite{Xie2020},\cite{Zhou2020}\\
\midrule
\multicolumn{2}{c}{Validation/Testing} \\
\cmidrule(lr){1-2}\\
Usability Tests & \cite{Cheng2019},\cite{Eiband2018},\cite{Holzinger2020},\cite{Mualla2020},\cite{Naiseh2021},\\
& \cite{Tsai2021B},\cite{Zhou2020}
\\
A/B Tests & \cite{Mualla2020},\cite{Wolf2019}\\
Interviews & \cite{Bussone2015},\cite{Eiband2018},\cite{Mohseni2021},\cite{Naiseh2021},\cite{Rebanal2021},\cite{Brusilovsky2019},\cite{Tsai2021},\cite{Zhou2020}\\
Mental Models & \cite{Kulesza2010},\cite{Tsai2021B},\cite{Eiband2018},\cite{Mohseni2021},\cite{Tsai2021}\\
Questionnaires & \cite{Cheng2019},\cite{Mualla2020},\cite{Tsai2021B},\cite{Omeiza2021},\cite{Eiband2018}\\
\bottomrule
\end{tabular}%
\end{table}

\subsubsection{Requirements Engineering} 

\paragraph{Vision Definition: } Before requirements can be specified, broader visions are developed. A vision can refer to the capabilities, features, or quality aspects of a system \cite{Schneider2019}. A vision defines the long term goals of a project and can facilitate communication and the scope definition of a software project, boosting the chances of producing a successful system.  By building a shared vision, it is possible to resolve misunderstandings with respect to the system-related goals and stakeholders' expectations, and have a grip on the essential quality aspects that must be considered in the system \cite{Schneider2019,Glinz2015}. In this phase, the first thing that needs to be determined is to what extent explainability really provides additional value for stakeholders. After all, explainability should only be considered if it is identified as a need \cite{Chazette2020} and if it aggregates value to the system since in some cases the cost of explanations might outweigh their benefits~\cite{Bunt2012}. Qualitative practices can be used to support this process such as interviews and workshops. These practices help to understand aspects such as behaviors, attitudes, and domain-specific aspects such as technical, business, and environmental contexts. Understanding these aspects is fundamental to identify and specify requirements. We consider that vision definition is an important part of every software project, and that it occurs either intentionally or unintentionally. During this activity, the need for explainability may be assessed, and from there, it can be determined whether explainability is a non-functional requirement for the system, after which the other activities that we recommend can be carried out.

\paragraph{Stakeholder Analysis: } 
Langer et al.  \cite{Langer2021} discuss the importance of paying special attention to understanding the existing stakeholder groups in the case of explainable systems. Examples of stakeholder groups are non-technical end users, domain experts, IT experts, regulators, ethicists, supervisors, and customers. Identifying stakeholder groups is important to analyze which interests and needs the relevant stakeholder groups have with respect to explainability \cite{Nuseibeh2000}. Different stakeholders have different goals and needs regarding the software system, having also other requirements on explanations. Again, qualitative approaches such as interviews and end user observation give valuable information on users' genuine needs and expectations regarding the system, on where explainability  might be needed in the system, and on how explanations should be designed \cite{Brusilovsky2019,Eiband2018}. The most frequent practices used to achieve these goals are interviews, personas, scenarios, questionnaires, focus groups, and workshops. After the two first activities (vision definition and stakeholder analysis), broad explainability goals may be set, which can be further refined over the subsequent phases.

\paragraph{Back-End Analysis: }

In this activity, the algorithm's logic is evaluated or planned with respect to the explainability goals. The first stage of this activity is to determine if the component of the system to be explained already exists (e.g., explainability must be implemented into an existing system) or whether the algorithm must still be developed/integrated (e.g., a system developed from scratch). If it already exists, the first thing to consider is whether the algorithm to be explained (we will refer to it as the \enquote{back-end algorithm} from now on) is interpretable  \cite{Longo2020}. As an example, consider that the back-end algorithm is based on a machine learning model. Machine learning models, especially deep learning, can produce accurate system outputs but are often referred to as black-boxes, since their rationale cannot be easily understood \cite{Arrieta2020}. Even data scientists frequently struggle to grasp how the model produces its outcomes \cite{Melis2021}. In the machine learning domain, a model is considered \emph{interpretable} when it is possible to determine why it produces a specific outcome. In a nutshell, the more interpretable a model is, the easier it is to explain its rationale and outcomes. In some cases, the back-end algorithm consists of a model that is not interpretable, so that alternative explainability techniques (e.g., post-hoc explanations) need to be employed to explain it~\cite{Arrieta2020}.

Another factor to be analyzed in this case is whether the desired explanations require global or local interpretability. Global interpretability often relates to understanding how a model works in general, whereas local interpretability refers to explaining each specific prediction \cite{Kopitar2019}. Therefore, it is essential to examine and specify which explainability approach is required and practicable in light of the algorithm to be explained \cite{Longo2020}. Communication between team members is essential in order to understand if there are technical constraints that limit the attainment of the explainability goals, and to find appropriate solutions \cite{Liao2020}. Suitable practices are workshops, brainstorming sessions and interviews with team members.

\paragraph{Trade-off Analysis: }During trade-off analysis, it is important to evaluate how explainability interacts with other quality aspects. Chazette et al. \cite{Chazette2021} investigated the interaction between explainability and other quality aspects and identified 57 quality aspects on which explainability can have both a positive and negative impact. For instance, in the case of usability, on the positive side, explanations can increase the ease of use of a system \cite{Nunes2017} or lead to a more efficient use \cite{Zhou2019}. On the negative side, explanations can overwhelm users with excessive information \cite{Tsai2019} and can impair the user interface design. This negative or positive impact depends, in the end, on the design decisions toward explainability. During this activity, knowledge catalogues are useful artifacts that can assist software developers in avoiding quality-related conflicts and determining the best techniques for achieving the intended quality outcomes \cite{Chazette2020}. Hence, practitioners should focus on design decisions and interactions in order to turn explainability into a positive catalyst for other crucial quality aspects in modern systems. Practices from back-end analysis are also recommended for this activity.%

\paragraph{Requirements and Mental Models}
To define requirements, it is important to compare and understand what the goals of the different stakeholders in the process are and how these goals point to what is expected from the system when it comes to the communication with the user~\cite{Mishra2008}. Mental models are often mentioned in the literature as a way to capture expectations regarding the understandability and explainability of a system~\cite{Kulesza2010,Mohseni2021,Tsai2021,Eiband2018,Tsai2021B}, supporting the definition of requirements. For instance, after back-end and trade-off analysis, an \emph{expert mental model} can be defined. An expert mental model represents the expected behavior of the system, based on the designers' (e.g., the experts) conception. Expert mental models can be used to capture how the system should be understood and how an explanation should help to understand the system. Practitioners can use expert mental models as a reference when defining or refining requirements. In addition, expert mental models can serve as reference later on, to be compared with the users' mental models, to evaluate the understandability of the system or the quality of the explanations. Essentially, mental models can help to define the right requirements, and to meet the right design decisions.

\subsubsection{Design / Implementation}
\paragraph{Explainability Design: } During this activity, software practitioners must make decisions about the specific characteristics of explanations based on the requirements. Important design aspects concerning explanations should be specified in this phase: whether explanations should be static or interactive; the language to be used in the explanations (e.g., technical, casual); when explanations should be presented (e.g., before or after an event); and how the information should be presented (e.g., audio, text, and other UI aspects) \cite{Longo2020,Nunes2017}. During explainability design, prototypes (low-fidelity and high-fidelity prototypes) are useful for presenting design ideas and various forms of explanations during explainability design, allowing for rapid visualization and discussion of design concepts before the actual implementation. Prototypes can be evaluated for effectiveness and to assess whether they meet the explainability requirements. Once design decisions have been made, their implementation can take place according to the company's development culture.



\subsubsection{Validation / Testing}

\paragraph{Evaluation: } The evaluation activity checks whether the system is explainable, that is, whether the explanations are adequate or need to be improved. The focus of this activity is on end-user feedback in combination with prototype or version testing (A/B tests), since the effectiveness and quality of an explanation is subjective \cite{Martin2021}. 

Evaluating the explainability of a system is challenging, as each individual has different cognitive processes while understanding something \cite{Martin2021}. Since explanations fill knowledge gaps, understanding the end users' cognitive processes is essential. Therefore, an important aspect for the evaluation is, again, the concept of mental models, since they provide a good way to capture cognitive processes \cite{Eiband2018}. During the system evaluation, users' mental models can be compared to the expert mental model created during requirements engineering. Deviations of the end-user model from the expert model reveal misunderstandings and/or problems with explanation requirements that need to be addressed, by adapting the system itself or the provided explanations so that the system can be better understood~\cite{Eiband2018,Mohseni2021}.





\subsection{Interview Study}

In order to evaluate the insights of our literature review, we conducted an interview study. To understand in what context the study participants worked, the first point of the interview was to ask them what kind of software process is applied in their companies. All participants stated that they use an agile software development process: The smaller companies use Scrum, the larger ones apply the V-model extended with SAFe.
In the end of each iteration, there is a timeframe for testing or evaluation. This feedback loop was often emphasized by the participants as being essential.

\begin{figure}[!ht]
	\centering
	\includegraphics[width=\linewidth]{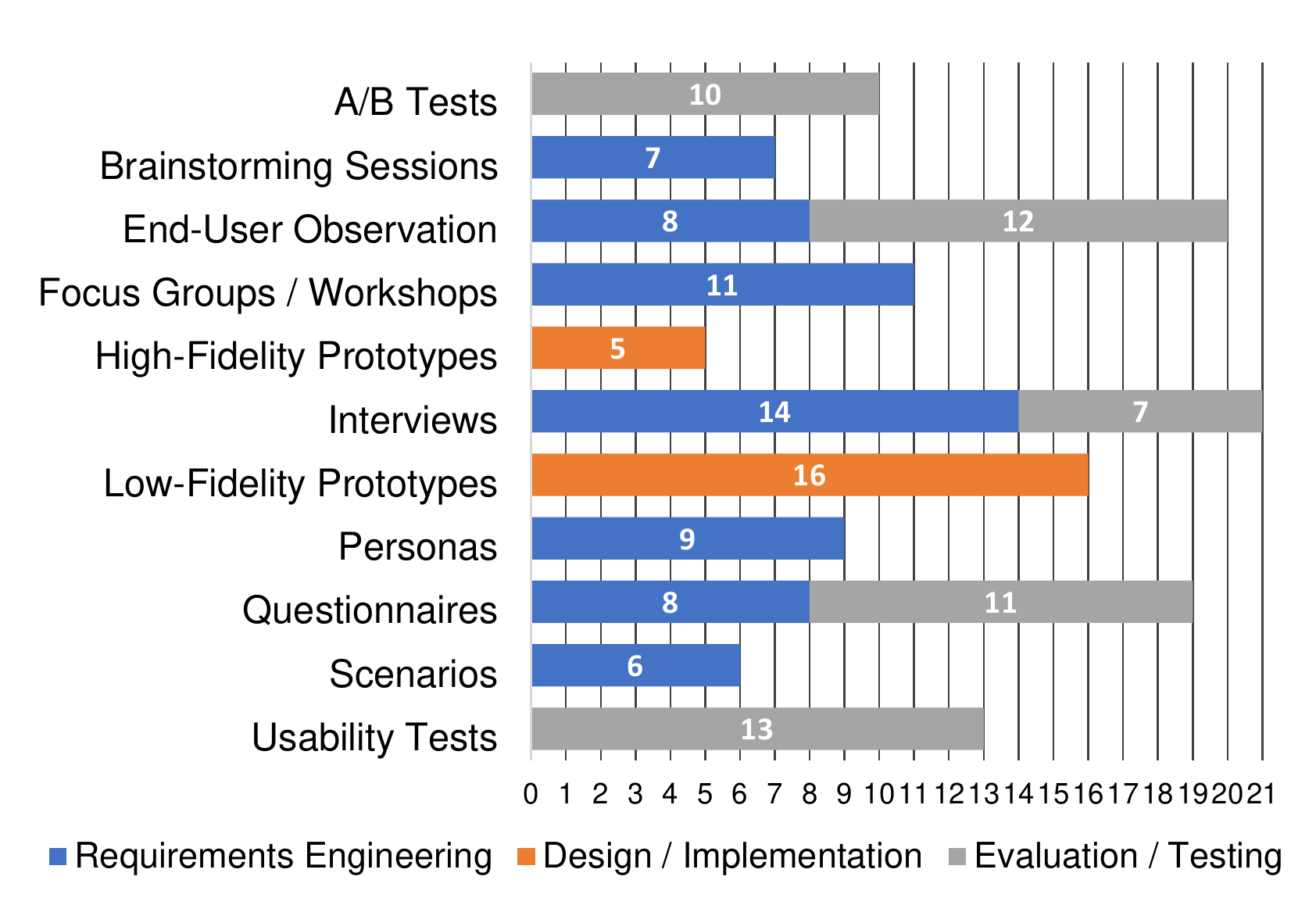}
	\caption{Recommended practices per phase (n=19)}
	\label{fig:InterviewResults}
\end{figure}

We compared the processes developed by our participants to the findings of the literature to validate the conclusions of the literature review. Figure \ref{fig:InterviewResults} shows which practices (and in which phase) are perceived as useful to support the development of explainable systems, according to the participants. 

Regarding the methods that can be used during requirements engineering, participants suggested interviews (73.7\%), focus groups and workshops (57.9\%), personas (47.4\%), questionnaires (42.1\%), brainstorming with customers and colleagues (36.8\%), scenarios (31.6\%), and/or end user observation (42.1\%). During design, participants suggested low-fidelity prototypes (84.2\%) (such as mock-ups and paper prototypes) and/or high-fidelity prototypes (26.3\%). Prototypes were often mentioned as a practice useful for requirements prioritization. For the evaluation, usability tests (68.4\%), end user observation (63.1\%), interviews (36.8\%), questionnaires (57.9\%) and/or A/B tests (52.6\%) were suggested.


Overall, the participants' recommendations match the findings in the literature. Even though the participants did not mention mental models explicitly, these were implicitly present in their statements at various points when emphasizing the need for a comparison between users' expectations and the system's functioning. Therefore, we see it as a useful and plausible practice. In addition, the interviewees recommended practices such as brainstorming sessions with the team and/or with stakeholders, and also end user observation. 

In the following sections, we discuss main points that were deemed important by the interviewees in addition to the activities or practices. These points include the significance of an iterative and user-centered approach, the impact of process type, how realistic the activities and practices are, and the potential problems of incorporating them into practice.

\subsubsection{Iterative and User-Centered Approach}

In summary, the participants were unanimously in favor of developing explainable systems using an iterative process. However, note that all participants work in an agile environment and might not be experienced with plan-driven approaches, which might have influenced their recommendation. All participants recommend an agile development process, since [sic] \enquote{agile procedures are iterative and allow changes to be made in an uncomplicated way}. Since explainability is still quite unknown and, hence, not explicitly addressed in existing software processes, participants advocate for a trial-and-error iterative procedure to develop explainable systems. This way, the users' understanding can be constantly evaluated and the design can be changed to improve the explainability of the system

One participant mentioned that the development of explainable systems is rather independent of the software development process and that it is also possible within a waterfall development process, but it can take a longer development time compared to agile approaches. Another participant affirmed that the waterfall development process could be less suitable for developing explainable systems due to its sequential nature since a perfect solution is not likely to be found upfront. One participant highlighted the importance of feedback loops in the process, since they allow to adapt the design according to user feedback. 

An end-user-centered process was recommended by all participants for the development of explainable systems. 
Since all participants stated that the development of explainable systems would be possible using the current development methodology in their companies, and often used the process as a base for their recommendations, we asked the participants if they see any need for optimization. All participants agreed that adopting more end-user-centered practices can help to better capture end-user requirements and evaluate design decisions. Eight participants mentioned that including the team members in the evaluation phase and making them aware of user feedback can help them better understand the existing challenges and design concerns.

\subsubsection{Realism and Challenges}

Four participants mentioned that the best development process for explainable systems would be useless if the team is not aware of the importance of explainability. Therefore, the motivation for integrating explainability should be clear to the team and, at best, the team should be convinced of and inspired by its value. Just as the requirement for, e.g., security is constantly considered, explainability must also be kept in mind whenever changes or adjustments are made. 

Finally, the participants were asked about the suitability of their recommendations when considering their company's context. As an example, resources such as time, cost, or the number of employees in a company were mentioned by the interviewer as possible factors influencing the practicability of the suggested activities and practices. All participants agreed that the activities and practices could be integrated. Three participants noted that the choice of the activities and practices depends on each company's resources, on the team, and on the product. Seven participants said that a trial-and-error approach is necessary until an ideal strategy about the optimal activities and practices, and how to integrate them into the company's specific context is identified.

\section{Discussion}\label{sec:discussion}


New quality requirements such as explainability always bring new challenges and uncertainties as how to proceed or adapt the implementation of systems. Therefore, we combined the findings of a literature review with the perspectives of practitioners gathered through an interview study to compile a set of activities and practices for the development of explainable systems. Practitioners believe that existing user-centered approaches and practices are effective when dealing with explainability in a software project, which supports the literature findings. The participants also expressed familiarity with these practices. Based on this analysis, we answer our first research question:

\par
\noindent \paragraph{$\rightarrow$ \textbf{Answer to RQ1:}} We identified six core activities that should be considered when developing explainable software systems:  vision definition, stakeholder analysis, back-end analysis, trade-off analysis, explainability design, and evaluation. These activities can be supported by user-centered practices such as end user observation, interviews, questionnaires, personas, prototypes, focus groups, workshops, brainstorming sessions, scenarios, storyboards, A/B and usability tests.
\par
\bigskip 
This finding supports the conclusions of a previous work. In this work, Chazette and Schneider~\cite{Chazette2020} recommend the adoption of existing user-centered practices during requirements engineering to develop explainable systems. One the one hand, this approach is derived from an empirical study with end users, in which the participants stated their opinions on potential issues regarding embedded explanations. The responses were analyzed and compared to existing usability heuristics. The heuristics were shown to be helpful in the resolution of the pointed issues and to support the design of adequate explanations which, in turn, favors the assumption that user-centered methods are adequate to develop explainable systems. On the other hand, this assumption had not yet been validated beyond this study, and, more importantly, the feedback of software experts on whether this approach is compatible with the reality of the industry was still missing. As a result of the current investigation, we discovered that these user-centered practices are also used in the literature.
Additionally, feedback from software experts confirms that these practices are also feasible in the industry, indicating an alignment between research and practice. 
 

Our study lays the foundation on the necessary activities that should be integrated into the software lifecycle to assist practitioners through the steps required for the development of meaningful, explainable systems. The set of activities proposed in this paper and the respective practices constitute our recommendation on how to develop explainable systems: 

\noindent \paragraph{$\rightarrow$ \textbf{Answer to RQ2:}} We synthesized the  practices identified in the literature review in six core activities that belong to three lifecycle phases: requirements engineering, design/implementation, and validation/testing. In order to develop explainable systems, these activities need to be included into the software lifecycle and therefore carried out in the real process.
\par

\bigskip 
Based on the answers of our two research questions, we consider the main takeaways of our work:

\begin{enumerate}
    \item End-user-centered practices are most suited to understand explainability requirements and develop explainable systems.
    \item The six core activities can be integrated into existing processes, allowing them to be carried out in both traditional and agile settings.
\end{enumerate}





These takeaways evidence the flexibility of the set of activities and practices recommended in this paper. Firstly, our results point to the use of practices rather than methods. This goes in line with findings from recent research on software development processes showing that practices appear to be way more important than methods, as the use of practices does not depend on the selected development process~\cite{tell2019hybrid,klunder2020determining}, making the activities and practices applicable in either waterfall or agile development environments.

Secondly, there is evidence that many of the methods and practices coming from the research community have not been embraced by practitioners, since it is often difficult for practitioners to find a way to incorporate new research ideas into their busy workdays \cite{Ameller2012}. The major concern of software professionals is cost, both in terms of time and money \cite{Kluender2019}. As a consequence, embracing current activities and practices rather than introducing new ones that increase costs is a favorable option~\cite{Chazette2020}. Even though  integrating such activities and practices might still represent a not negligible overhead for the industry, the proposed activities can still be adapted according to the necessity of each particular project. 

It is important to notice that the activities cover the software lifecycle only partially, addressing only requirements engineering, design, and testing. We consider that current knowledge on how to structure the remainder of the development process is equally true for the creation of explainable systems and, hence, is outside the focus of this study. However, even though the activities can be integrated into an existing process alongside other activities and phases, we do not discard the need for further research to cover the whole lifecycle. In either case, it was possible to identify that there is a special focus on the requirements process and on validating these requirements, both in the works found in the literature and in the interviews. Therefore, requirements-related activities in the process should be given special attention since the development of explainable systems is heavily reliant on meticulous requirement analysis, as well as how those requirements are implemented and transformed into explanations, and how those explanations are displayed on the user interface.

 


\section{Conclusion}\label{sec:conclusion}

Although explainability is considered a catalyst for essential qualities in modern software systems such as transparency and fairness, it is not yet clear how to approach it in practice. We conducted a literature review to understand how explainability is addressed and which activities and practices can be used to design explanations. Based on the results from the literature, we built a recommendation consisting of six core activities and associated practices for the development of explainable systems. We conducted an interview study with 19 software practitioners to assess the feasibility of the activities in practice.

As researchers, we have to take action in investigating methods and techniques that are aligned with the practice and offer advantages for practitioners instead of more overhead. Therefore, we recommend a set of activities and well-known practices that may be included in the development process, instead of conceiving a completely new process with unfamiliar practices for the development of explainable systems. Our research shows that software practitioners can employ well-known user-centered practices to elicit, implement, and test requirements that are aligned with users' needs, as well as their demands and context. 

Future research is needed to learn more about the other phases of the software lifecycle and whether they need to be adapted. So far, despite the attempts to analyze software processes, there were no proposals that explicitly focus on the development of explainable systems. Therefore, we believe our proposal provides both a starting point and a foundation that can be further updated and improved.


\begin{acks}
This work was supported by the research initiative Mobilise between the Technical University of Braunschweig and Leibniz University Hannover, funded by the Ministry for Science and Culture of Lower Saxony. We thank our colleague Nils Prenner for his valuable feedback and fruitful discussions.
\end{acks}

\bibliographystyle{ACM-Reference-Format}
\bibliography{sample-base}

\appendix

\end{document}